\documentclass[journal]{IEEEtran}
\usepackage{cite}

\ifCLASSINFOpdf
   \usepackage[pdftex]{graphicx}
\else
   \usepackage[dvips]{graphicx}
\fi
\usepackage{amsmath}
\ifCLASSOPTIONcompsoc
  \usepackage[caption=false,font=normalsize,labelfont=sf,textfont=sf]{subfig}
\else
  \usepackage[caption=false,font=footnotesize]{subfig}
\fi
 \let\MYorigsubfloat\subfloat
 \renewcommand{\subfloat}[2][\relax]{\MYorigsubfloat[]{#2}}
\usepackage{bm}
\usepackage{multirow}
\allowdisplaybreaks

\usepackage{lineno}
\usepackage{booktabs,multirow,amsthm,amsmath,mathrsfs}
\usepackage{algorithm}
\usepackage{algorithmic} 
\usepackage{amssymb} 

\usepackage{changepage}

\usepackage{color}

\begin{document}
%

\title{\Huge{Q-SR: An Extensible Optimization Framework for Segment Routing}}
%
%
%

\author{Jianwei~Zhang
	\thanks{This work was supported by the National Natural Science Foundation of China under Grant No. 61902346.
		\textit{(Corresponding author: Jianwei Zhang.)}
	}
	\thanks{Jianwei Zhang (janyway@outlook.com) is with the 
		School of Computer and Computing Science, 
		Zhejiang University City College, Hangzhou, China.}
}

\markboth{MANUSCRIPT, 2020}%
{Shell \MakeLowercase{\textit{et al.}}: Bare Demo of IEEEtran.cls for IEEE Journals}
%



\maketitle

\begin{abstract}
Segment routing (SR) combines the advantages of source routing supported by centralized software-defined networking (SDN) paradigm and hop-by-hop routing applied in distributed IP network infrastructure.
However, because of the computation inefficiency, it is nearly impossible to evaluate whether various types of networks will benefit from the SR with multiple segments using conventional approaches.
In this paper,
we propose a flexible $Q$-SR model as well as its formulation in order to fully explore the potential of SR from an algorithmic perspective. 
The model leads to a highly extensible framework to design and evaluate algorithms that can be adapted to various network topologies and traffic matrices.
For the offline setting, we develop a fully polynomial time approximation scheme (FPTAS) which can finds a $(1+\omega)$-approximation solution for any specified $\omega>0$ in time that is a polynomial function of the network size.
To the best of our knowledge, the proposed FPTAS is the first algorithm that can compute arbitrarily accurate solution.
For the online setting, we develop an online primal-dual algorithm that proves $O(1)$-competitive and violates link capacities by a factor of $O(\log n)$, where $n$ is the node number.
We also prove performance bounds for the proposed algorithms. 
We conduct simulations on realistic topologies to validate SR parameters and algorithmic parameters in both offline and online scenarios.
\end{abstract}


\IEEEpeerreviewmaketitle

\section{Introduction}
Segment routing (SR) combines the advantages of source routing supported by centralized software-defined networking (SDN) paradigm and hop-by-hop routing applied in distributed IP network infrastructure \cite{sr_survey,sr_survey2,Incremental}.
The key idea behind SR is to break a routing path into multiple segments using a sequence of SR-node (a.k.a. intermediate node) in order to control the routing path more flexibly and hence improve network utilization. 

In parallel with the SR, middleboxes have become ubiquitous in SDN as well as data center networks (DCN).
Service function chaining (SFC) is a set of operations to steer traffic through an ordered list of physical or virtual middleboxes
which provide network functions such as VPN, NAT, DPI, and firewall.
%
From another perspective, SR can be viewed as the supporting technology of a large variety of novel network technologies at the network layer, such as SFC \cite{srVNF}, network function virtualization (NFV) \cite{sr_cloud}, 5G \cite{5g}, SD-WAN \cite{sr_survey2}.
They share a common technical ground on the modeling and algorithm design from the multi-commodity flow (MCF) theory, while having disparate orientations. 

%
%
Existing literature claim that 2-SR (SR using at most 2 segments) can achieve near-optimal performance \cite{bhatia2015sr,computation-efficient,schuller2018,schuller2020}.
Thus, the routing paths between the source and target nodes usually take on a short and wide shape, 
i.e., there are usually a large number of 2-hop paths; here each hop means a shortest path routing.
This is quite different from the case in SFC, where the routing paths are often long and narrow.
\textit{Due to the computation inefficiency, it is nearly impossible to evaluate whether various types of networks will benefit from the SR with multiple segments using conventional approaches \cite{schuller2018}.}
To this end, in this paper, we try to address the following challenges in SR networks.
\begin{itemize}
	\item How to establish a flexible and extensible model to optimize network throughput by leveraging SR with multiple segments?
	\item How to design efficient algorithms to solve the model in both offline and online scenarios?
\end{itemize}	

To tackle the above challenges, we aim to fully explore the potential of SR from an algorithmic perspective.
Similar to the methodology in \cite{jadin2019cg4sr}, we will not consider practical hardware (e.g. routers) or software (e.g. protocols) limits on the segment number.
Specifically, the main contributions of this paper are the following:
\begin{itemize}
	\item We propose a flexible $Q$-SR model as well as its formulation where segment number, SR-node number, intra-segment routing policy are all parameterized. 
	The model leads to a highly extensible framework to design and evaluate algorithms that can be adapted to various network topologies and traffic matrices.
	\item For the offline setting, we develop an \textit{fully polynomial time approximation scheme} (FPTAS) which can finds a $(1+\omega)$-approximation solution for any specified $\omega>0$ in time that is a polynomial function of the problem size.
	The proposed FPTAS is the first algorithm that can compute arbitrarily accurate solution and only on this basis can we further evaluate whether using multiple segments is inevitable on various types of networks.
	\item For the online setting, we develop an online primal-dual algorithm that proves $O(1)$-competitive and violates link capacities by a factor of $O(\log n)$, where $n$ is the node number.
	\item We prove performance bounds for the proposed algorithms. 
	We conduct simulations on realistic topologies to validate SR related parameters and algorithmic parameters in both offline and online settings.
\end{itemize}

The rest of this paper is organized as follows. 
We review related work in Section II. 
We introduce the system model and preliminaries in Section III.
We formulate the offline and online network throughput maximization problems for SR and develop approximation and online algorithms in Sections IV and V, respectively.
The min-cost SR-path computation module is presented in Section VI.
We present simulation results in Section VII. 
Finally, we discuss important extentions in Section VIII and conclude in Section IX. 
All proofs are presented in the appendix. 
Main notation is summarized in Table~\ref{tb:notation}.

\section{Related Work}\label{sec:relatedwork}
\subsection{Segment Routing Using Multiple Segments}

Bhatia \textit{et al.} \cite{bhatia2015sr} for the first time formulate a generic SRTE problem to minimize maximum link utilization, where all intermediate nodes are used to construct optimal segment routing paths and they also propose a 2-SR online algorithm.

To jointly optimize the efficiency of intermediate nodes selection and the subsequent flow assignment, Settawatcharawanit \textit{et al.} \cite{computation-efficient} formulate a bi-objective mixed-integer nonlinear program (BOMINLP) to investigate the trade-off between link utilization and computation time.
They conclude that the maximum link utilization performance of 2-SR is indentical to $\infty$-SR.
Thus, they focus on limiting the candidate paths lengths as well as reducing the computation overheads by a \textit{stretch bounding} method.
%


Pereira \textit{et al.} \cite{pereira2020threee} propose a single adjacency label path segment routing (SALP-SR) model that forwards traffic flows using at the most three segments.
They also propose an evolutionary computation approach to improve traffic distribution. 
As applications, they also extend the model to handle semi-oblivious traffic matrices and address link failures. 

Jadin \textit{et al.} \cite{jadin2019cg4sr} formulate the SRTE problem into an ILP, and propose a CG4SR approach that combines the column generation and dynamic program techniques.
The approach can only obtain near optimal solutions with gap guarantees by realistic experiments.
They also compute a stronger lower-bound than traditional MCF approach through experiments.

Li \textit{et al.} \cite{sr_milp} find that SR without support of adjacency segments cannot reach the optimum.
To fully support adjacency segments in SR, they propose an extended LP formulation for 2-SR and an MILP formulation for $K$-SR.
Due to the computation complexity, the MILP is further simplified to prevent excessive flow splitting or using long paths. 

SR combines the advantages of centralized inter-segment source routing and distributed intra-segment hop-by-hop routing.
Unlike the above works, the proposed $Q$-SR model in this paper provides the maximum freedom to deploy SR-nodes and predict the performance.
Our algorithm proves a $(1+\omega)$-approximation solution, that is, it can be arbitrarily close to the optimum.
In our simulation, the proposed algorithms can even support rapid computation for the number of segments as large as $O(n)$.

\subsection{Service Function Chain}
Starting from the classical MCF model, Cao \textit{et al.} \cite{steering} consider the policy-aware routing problem in both offline and online settings, where a traffic demand must traverse a predetermined ordered list of middleboxes.
However, the only resource constraint is put on link capacities; the middleboxes do not consume any resources. 

Further, Charikar \textit{et al.} \cite{in_network_processing} propose a new kind of MCF problem, where the traffic flows consume bandwidth on the links as well as processing resources on the nodes.
They also formulate the problem via an LP model and develop an efficient combinatorial algorithm to solve the model approximatedly with arbitrary accuracy.

Recently, 
more realistic SFC models for unicast \cite{anu_unicast} and multicast \cite{anu_multicast}, which incorporate link bandwidth, residual energy in mobile devices and cloudlet computing capacity constraints in the context of NFV-enabled MEC networks are proposed.
Notice that these models assume that the network elements (APs or cloudlets) in the backbone are connected via wireline links.

Although SR and SFC belong to different areas of research, they are very close in terms of the MCF-based models and algorithms.
In SR, all the available resources, including network links and SR-nodes, should be jointly managed to optimize the TE objectives at the network layer \cite{moreno2017,schuller2018}.
While in SFC, the middleboxes impose extra computation and storage constraints on a wider range of objectives from network layer to application layer.
Therefore, the two areas can borrow ideas and merit from each other. 

\begin{table}[!t]\small
	\caption{\label{tb:notation}Notations}
	\centering
	\begin{tabular}{p{2cm}p{6cm}}
		\toprule
		\textbf{Notation} & \textbf{Description} \\
		\midrule
		$G=(V, E)$ & SR network $G$, where $V$ is the node set and $E$ the link set. \\
		${{\widetilde{G}}_{r}}=({{\widetilde{V}}_{r}},{{\widetilde{E}}_{r}})$ & Auxiliary graph constructed for request $r$. \\
		$m, n$ & Node number and link number of $G$. \\
		$c_e$ & Capacity of link $e$. \\				
		$r$, $s_r$, $t_r$ & Request $r$, its source node and target node. \\
		$d_r$ & Size of request $r$. \\
		$K_r$ & Set of all possible SR-node lists for request $r$. \\
		$N_r$ & Available SR-nodes for request $r$. \\
		$Q_r$ & Maximum number of segments for request $r$. \\
		${P_r^k}$ & SR-path via SR-node list $k$ due to request $r$. \\
		${g_{r}^{k}(e)}$ &  Mapping coefficient from ${P_r^k}$ to link $e$, i.e. the amount of flow routed on link $e$ through SR-node list $k$ due to a unit request $r$. \\	
		${X_{r}^{k}}$ & Fraction of request $r$ routed through SR-node list $k$. \\
		${x_{r}^{k}}$ & Flow amount of request $r$ routed through SR-node list $k$. \\
		$l_e$ & Dual variable associated with each link $e$. \\
		$z_r$ & Dual variable associated with each request $r$. \\
		$\lambda$ & Multiplier that request size $\lambda {{d}_{r}}$ can be supported for $r$. \\
		$\epsilon$ & Tunable parameter of FPTAS. \\
		$\phi$ & Tunable parameter of online algorithm. \\
		SR & Segment routing. \\
		LP & Linear program. \\
		MCF & Multi-commodity flow. \\
		ECMP & Equal-cost multipath. \\
		MF & Middlebox fabric. \\
		\bottomrule
	\end{tabular}
\end{table}

\section{System Model}
We model an SR network with a directed graph $G=(V, E)$, where $V$ represents the node set and $E$ the link (edge) set.
The number of nodes and links are denoted by $n$ and $m$, respectively.
The network is not necessarily assumed symmetric, i.e., some links may not be bi-directional.

We introduce the following SR parameters in the \textit{$Q$-SR framework}, which can be illustrated by Fig.~\ref{fg:agc} in Section \ref{sec:kernel}.
\begin{itemize}
	\item \text{SR-node set} (a.k.a. the \textit{width} of MF in Fig.~\ref{fg:agc});
	\item \text{Segment number} (a.k.a. the \textit{length} of MF in Fig.~\ref{fg:agc});
	\item \text{Intra-segment routing};
	\item \text{Inter-segment routing}.
\end{itemize}

Given a request $r$, assume the set of available SR-nodes is $N_r$ and
the (maximum allowable) segment number is $Q_r$, then $Q_r-1 \le |N_r|$.

By definition, $Q$-SR may even degenerate to 1-SR, i.e. the shortest path routing,
or generalize to $\infty$-SR, i.e. the MCF routing,
which can use all simple paths available to achieve the highest performance in theory while suffering from the largest cost.

Define ${{K}_{r}}$ as the set of all possible SR-node lists, i.e.:
\[{{K}_{r}}=\left\{ ({{k}_{1}},{{k}_{2}},...,{{k}_{{{Q}_{r}}-1}}):{{k}_{i}}\in {{N}_{r}}\cup \varnothing ,\forall i \right\}\]

The framework is flexible and extensible to support innovations in SR.
For instance, under this framework, both node segment and adjacency segment in SR can be supported.
In this paper, we restrict our attention to inter-segment routing, assuming that the intra-segment routing multpaths are predetermined using some link-state routing protocols.


%
%
%

\subsection{SR-Function for Intra-Segment Routing}
Let ${f_{uv}}(e)$ represent the flow on link $e$ when unit flow is routed from $u$ to $v$ according to some link-state routing policy. 
The routing policy may be the shortest path algorithm (ECMP permitted), DEFT \cite{deft}, PEFT \cite{peft}, and etc. 
Note that ${f_{uv}}(e)$ is uniquely determined by the IGP link weights, which have no relation with the link length system $l$ in the dual problem. 
The link-state based routing policy based on the IGP link weights.
Therefore, we treat ${f_{uv}}(e)$ as input parameters of the solution algorithms.
The definition and computation of ${f_{uv}(e)}$ consider the hierarchical structure of the Internet.
Unlike previous works, we need not to compute ${f_{uv}(e)}$ over the entire node set $N$.
This provides more operational flexibility of the SR-nodes deployment.

Given an SR-node list $k \in K_r$ for request $r$ and the notations ${{k}_{0}}:=s_r$ and ${{k}_{{{Q}_{r}}}}:=t_r$.
Define the \textit{SR-function} $g_r^k(e)$ as
\[g_{r}^{k}(e)=\sum\limits_{t=0}^{{{Q}_{r}}-1}{{{f}_{{{k}_{t}}{{k}_{t+1}}}}(e)}.\]

Therefore, $g_r^k(e)$ calculates the flow on link $e$ if a unit flow is routed from $s_r$ to $t_r$ through SR-node list $k$. 
It is predetermined by the network topology, link weights and intra-segment routing policy including but not limited to the shortest path policy.  
Note that if the equal-cost multipath (ECMP) routing is employed, then $g_r^k(e)$ can be fractional 
and that if there is a link traversed more than once, $g_r^k(e)$ can be larger than one.
Thus, the path, referred to as \textit{SR-path}, is a \textit{generic path} rather than a \textit{simple path}.
In the following sections we will see that the SR-path constitutes the meta-structure of the proposed algorithms.
The examples in the next section illustrate how traffic will be split across the SR-path. 

It is not hard to see that \textit{adjacency segments} \cite{jadin2019cg4sr,pereira2020threee} can also be supported in the $Q$-SR mode as well as the proposed algorithms in the following sections.
For instance, suppose link $uv$ is an adjacency segment, we only need to make a simple assignment ${f_{uv}}(uv)=1$.
For clarity and without loss of generality, we focus on the \textit{node segments} in this paper.
%
%

\subsection{Illustrative Examples}
\begin{figure}[!h]
	\centering
	\includegraphics[angle=0, width=0.43\textwidth]{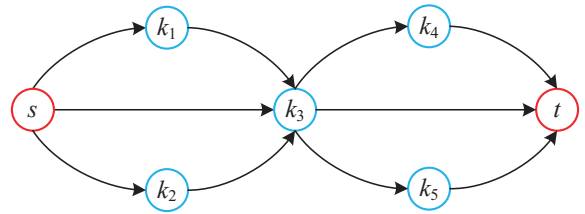}
	\caption{Inefficiency of 2-SR (Example 1). All links have identical capacities 100 and weights 1.}
	\label{fg:2sr}
\end{figure}

\textit{Example 1:} The close-to-optimal performance of the 2-SR setting, when being applied to real networks, has been claimed in many literature, e.g. \cite{bhatia2015sr,computation-efficient,schuller2018}.
In \cite{schuller2018}, an unrealistic topology is constructed to validate this point.
Here we give another counter example to illustrate the inefficiency of 2-SR, see Fig.~\ref{fg:2sr}.
The topology we used here, however, can be seen as a highly abstract multi-domain network structure.
Suppose all links have identical capacities 100 and weights 1.
Under the 2-SR setting, the maximum throughput from $s$ to $t$ is 100 even though all nodes in blue color are selected as candidate SR-nodes.
This is due to the fact that some paths, e.g., $(s,k_1,k_3,k_4,t)$, cannot be utilized.
In the 3-SR setting, however, this path can be activated if $k_1$ and $k_4$ are chosen as SR-nodes.
Similarly, all paths from $s$ to $t$ become available in this setting, thereby achieving the maximum thoughput 300.
This type of topology is quite common in the case of inter-domain routing.
The severe inefficiency shown in this example originates from the misalignment of link weights setting and the objective of throughput maximization.
To tackle this issue, we should steer traffic to non-shortest paths using link-state based routing protocols, just as the well known DEFT \cite{deft}, PEFT \cite{peft}, and the method in \cite{pereira2020threee}.

Another point we want to emphasize is that the flexibility and resultant complexity of intra- and inter-segment routing can be converted into each other.
More exactly, the maximum throughput can also be achieved using only $k_3$ as the SR-node by appropriately setting the values of ${f_{sk_3}}(e)$ and ${f_{k_3t}}(e)$.
The theoretical analysis to this convertion is still an open challenge.

\begin{figure}[!h]
	\centering
	\includegraphics[angle=0, width=0.49\textwidth]{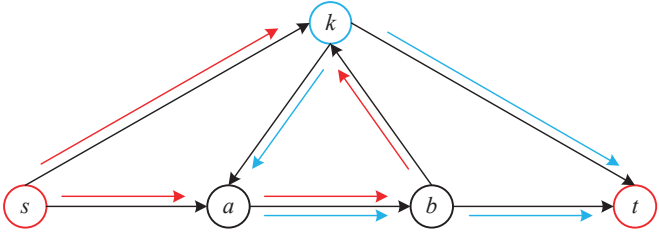}
	\caption{Intra-segment routing (Example 2). There is only one SR-node $k$.}
	\label{fg:multipath}
\end{figure}

\textit{Example 2:} 
In Fig.~\ref{fg:multipath}, we want to send a unit flow from $s$ to $t$ and there is only one SR-node $k$.
Suppose all the links have unlimited capacities and identical IGP weights 1.
According to the shortest path policy, only the paths (also links) $(s,k)$ and $(k,t)$ are utilized.
If the weights of links $sk$ and $kt$ are raised to 3, the ECMP policy could be activited.
If their weights are further raised to 5, only the paths $(s,a,b,k)$ and $(k,a,b,t)$ can be utilized.
In other words, in such case, no matter how inter-segment routing is optimized, links $sk$ and $kt$ can never be used.
To solve this problem, we only need to introduce the adjacency segments $sk$ and $kt$, i.e. setting ${f_{sk}}(sk)=1$ and ${f_{kt}}(kt)=1$. 

we emphasize again that in this paper we leave the intra-segment routing as an input parameter in the $Q$-SR framework and focus on the inter-segment routing optimization.

\section{Offline Network Throughput Maximization}

In the offline network throughput maximization problem, we know all the routing requests in advance.
The objective is to simultaneously maximize the throughput of all the requests.
We first present the offline formulation and thereby develop an approximation algorithm.

\subsection{Problem Formulation}
Based on the \textit{maximum concurrent flow problem} \cite{fptas2007}, the offline problem can be formulated as the following LP.
\begin{align}
(P_{\rm off})~\textbf{max}~&\lambda \nonumber\\
\textbf{s.t.}~
\label{eq:poff:f} &\sum\limits_{k \in {K_r}} {x_r^k}  \ge \lambda {d_r},&\forall r. \\
\label{eq:poff:c} &\sum\limits_r {\sum\limits_{k \in {K_r}} {g_r^k(e)x_r^k} }  \le {c_e},&\forall e. \\
&x_r^k \ge 0,&\forall r, \forall k \in {K_r}. \nonumber\\
&\lambda \in \mathbb{R}. \nonumber
\end{align}


Similar to a typical MCF formulation, constraints (\ref{eq:poff:f}) and (\ref{eq:poff:c}) imply the flow conservation and capacity limitation, respectively.
By associating a length $l_e$ with each link $e$ and a weight $z_r$ with each request $r$,
we can write the dual to the above LP as:
\begin{align}
(D_{\rm off})~\textbf{min}~&D(l): = \sum\limits_e {{l_e}{c_e}} \nonumber\\
\textbf{s.t.}~
\label{eq:poff:dual:l} &\sum\limits_e {g_r^k(e){l_e}}  \ge {z_r},&\forall r,\forall k \in {K_r}. \\
\label{eq:poff:dual:z} &\sum\limits_{r}{{{d}_{r}}{{z}_{r}}}\ge 1. \\
&z_r \ge 0,&\forall r. \nonumber\\
&{l_e} \ge 0,&\forall e. \nonumber
\end{align}

Using the definitions of $g_r^k(e)$, an alternate way to write the dual constraint is
\[\sum\limits_{t = 0}^{{Q_r} - 1} {\sum\limits_e {{f_{{k_t}{k_{t + 1}}}}(e){l_e}} }  \ge {z_r}\]

In the worst case, the SR-node list $k \in K_r$ can be as long as $O(n)$ and each position of $k$ can be empty or occupied by any SR-node.
Thus, the number of variables in $(P)$ can be at most $O(n^{n+2})$.
As far as we know, no tractable general-purpose LP solver can be directly applied to so large a problem. 

\subsection{Approximation Algorithm}

We design an FPTAS to solve the problem \cite{fptas2007,sr_restoration}.
The FPTAS is a primal-dual algorithm which includes an outer loop of a primal-dual update and an inner loop of min-cost SR-path computation. 

The algorithm to solve the problem starts by assigning a precomputed length of $\frac{\delta }{{{c_e}}}$ to all links $e$.

The algorithm proceeds in phases.
In each phase, we route $d_r$ units of flow from node $s_r$ to $t_r$, for each request $r$. 
A phase ends when all requests are routed.

The flow of value $d_r$ of request $r$ is routed from $s_r$ to $t_r$ in multiple iterations as follows.
In each iteration, a min-cost SR-path from $s_r$ to $T_r$ that minimizes the left-hand side of constraint (\ref{eq:poff:dual:l}) under current link lengths is computed.

The path is computated in Section \ref{sec:kernel}.
The bottleneck of this path, i.e. the maximum amount of flow that can be sent along this path, is given by
\[d_b=\underset{e\in P_{r}^{k}}{\mathop{\min }}\,\frac{{{c}_{e}}}{g_{r}^{k^*}(e)}\]

The amount of flow sent along this path in a step, denoted by $\Delta$, is also bounded by the remaining amount of flow for $r$, denoted by $d$, i.e.:
\[\Delta  = \min \left\{ {d,{d_b}} \right\}\]

%
%
%
%
%
%
%
%
%
%

After the flow of value $\Delta$ is sent through the SR-node list $k^*$, the flow value and the link length at each link along the path $P_{r}^{k^*}$ are updated as follows:

1) Update the flow values $f_e$ as
\[{f_e} \leftarrow {f_e} + g_r^{k^*}(e)\Delta\]

2) Update the link lengths $l_e$ as
\[{{l}_{e}}\leftarrow {{l}_{e}}\left( 1+\epsilon \frac{g_{r}^{k^*}(e)\Delta }{{{c}_{e}}} \right)\]

The update happens after each iteration associated with routing a portion of flow $d_r$. 
The algorithm terminates when the dual objective function value $D(l)$ becomes less than one.

When the algorithm terminates, dual feasibility constraints will be satisfied. 
However, link capacity constraint (\ref{eq:poff:c}) in the primal solution will be violated, since we were working with the original (not the residual) link capacities at each stage. 
To remedy this, we scale down the traffic at each link uniformly so that the link capacity constraints are satisfied.

\begin{algorithm}[!t]
	\caption{FPTAS for $Q$-SR Model}\label{alg_fptas}
	\begin{algorithmic}
		\STATE \textbf{Input:} ${l_e} \leftarrow \frac{\delta }{{{c_e}}},\forall e$; $f_e\leftarrow 0 ,\forall e$; $\rho\leftarrow 0$
		\STATE Initialize ${f_{uv}(e)},\forall u,v \in {N_r} \cup \{ {s_r},{t_r}\} ,\forall r$ via the IGP routing policy using primal link weight system.
		\WHILE{$D (l) < 1$}			
		\FOR{$\forall r \in R$}
		\STATE $d=d_r$
		\STATE Invoke \textit{Auxiliary Graph Construction}.
		\WHILE{$r>0$}
		\STATE Invoke \textit{Mincost Computation Module} to compute the optimal SR-node list $k^*$, and denote $P_{r}^{k^*}$ as the optimal SR-path.
		\STATE $d_b=\underset{e\in P_{r}^{k}}{\mathop{\min }}\,\frac{{{c}_{e}}}{g_{r}^{k^*}(e)}$
		\STATE $\Delta  = \min \left\{ {d,{d_b}} \right\}$
		\STATE $d=d-\Delta $
		\FOR{$\forall e\in P_r^{k^*}$}
		\STATE ${f_e} \leftarrow {f_e} + g_r^{k^*}(e)\Delta$
		\STATE ${{l}_{e}}\leftarrow {{l}_{e}}\left( 1+\epsilon \frac{g_{r}^{k^*}(e)\Delta }{{{c}_{e}}} \right)$
		\ENDFOR
		\ENDWHILE
		\ENDFOR
		\STATE $\rho \leftarrow \rho+1$
		\STATE $D (l) \leftarrow \sum\limits_e {{l_e}{c_e}}$
		\ENDWHILE
		\STATE ${f_e} = \frac{{{f_e}}}{{{{\log }_{1 + \epsilon }}\frac{{1}}{\delta }}},\forall e$
		\STATE $\lambda  = \frac{{\rho - 1}}{{{{\log }_{1 + \epsilon }}\frac{1}{\delta }}}$
		\STATE \textbf{Output:} ${f_e}$; $\lambda$
	\end{algorithmic}
\end{algorithm}

\textbf{Theorem 1:}
For any specified $\omega >0$, Algorithm \ref{alg_fptas} computes a $(1+\omega )$-approximation solution. 
If the algorithmic parameters are $\epsilon (\omega )=1-{{(1+\omega )}^{-\frac{1}{3}}}$ and $\delta (\omega )={{\left( \frac{1-\epsilon }{m} \right)}^{\frac{1}{\epsilon }}}$, 
the running time is $O\left( \frac{|R|\log |R|}{\epsilon }{{\log }_{1+\epsilon }}\frac{m}{1-\epsilon }{{{T_{{\rm{SR}}}}}} \right)$, 
where $T_{\rm{SR}}$ is the time required to compute a min-cost SR-path.

\textit{Proof:}
See Appendix.
\qed

\section{Online Network Throughput Maximization}

In the online network throughput maximization problem, the routing requests arrive one by one without the knowledge of future arrivals.
The objective is to accept as many requests as possible.
We first present the online formulation and thereby develop an online primal-dual algorithm.
\subsection{Problem Formulation}
Based on the  \textit{maximum multicommodity flow problem} \cite{fptas2007}, the online problem can be formulated as the following ILP.
\begin{align}
(P_{\rm on})~\textbf{max}~&\sum\limits_r {d_r\sum\limits_k {X_r^k} }\nonumber\\
\textbf{s.t.}~
\label{eq:pon:f} &\sum\limits_{k \in {K_r}} {X_r^k}  \le 1,&\forall r. \\
\label{eq:pon:c} &\sum\limits_r {{d_r}\sum\limits_{k \in {K_r}} {g_r^k(e)X_r^k} }  \le {c_e},&\forall e. \\
&X_r^k \in \{ 0,1\},&\forall r, \forall k \in {K_r}. \nonumber
\end{align}

Similar to the offline formulation, constraints (\ref{eq:pon:f}) and (\ref{eq:pon:c}) imply the flow conservation and capacity limitation, respectively.
We then consider the LP relaxation of this problem where $X_r^k \in \{ 0,1\}$ is relaxed to $X_r^k \ge 0$. 
Note that $X_r^k \le 1$ is already implied by constraint (\ref{eq:pon:f}).
By associating a length $l_e$ with each link $e$ and a weight $z_r$ with each request $r$,
we can write the dual to the above LP as:
\begin{align}
(D_{\rm on})~\textbf{min}~&\sum\limits_r {{z_r}}  + \sum\limits_e {{l_e}{c_e}} \nonumber\\
\textbf{s.t.}~
\label{eq:pon:dual:c} &{z_r} \ge {d_r}\left( {1 - \sum\limits_e {g_r^k(e){l_e}} } \right),&\forall r,\forall k \in {K_r}. \\
&z_r \ge 0,&\forall r. \nonumber\\
&{l_e} \ge 0,&\forall e. \nonumber
\end{align}

\begin{algorithm}[!t]
	\caption{Online Algorithm for $Q$-SR Model}\label{alg_online}
	\begin{algorithmic}
		\STATE \textbf{Input:} $l_e \leftarrow 0,\forall e$
		\WHILE{Upon the arrival of request $r$}	 
		\STATE Invoke \textit{Auxiliary Graph Construction}.
		\STATE Invoke \textit{Mincost Computation Module} to compute the optimal SR-node list $k^*$ and $L$, and denote $P_{r}^{k^*}$ as the optimal SR-path.
		\IF{$L>1$}
		\STATE Reject $r$;
		\ELSE
		\STATE Accept $r$;	
		\STATE $X_r^{{k^{\rm{*}}}} \leftarrow 1$ 
		\STATE ${z_r} \leftarrow {d_r}\left( {1 - L } \right)$
		\FOR{$\forall e\in P_r^{k^*}$}
		\STATE
		${l_e} \leftarrow {l_e}\left( {1 + \frac{{g_r^{k^*}(e){d_r}}}{{{c_e}}}} \right){\rm{ + }}\frac{\phi }{{Qn}}\frac{{g_r^{k^*}(e){d_r}}}{{{c_e}}}$,
		\STATE where $\phi >0$ and $Q = \mathop {\max }\limits_r {Q_r}$.
		\ENDFOR
		\ENDIF
		\ENDWHILE
		\STATE \textbf{Output:} $X_r^k$
	\end{algorithmic}
\end{algorithm}
\subsection{Online Algorithm}
We design an online primal-dual algorithm which includes an outer loop of a primal-dual update and an inner loop of min-cost SR-path computation. 

The algorithm to solve the problem starts by assigning a precomputed length of zero to all links.

The algorithm proceeds in iterations and each iteration corresponds to a request.
Upon the arrival of a new request $r$, we try to route $d_r$ units of flow from node $s_r$ to $t_r$, for each request $r$. 

In each iteration, a min-cost SR-path from $s_r$ to $t_r$ that maximizes the right-hand side of constraint (\ref{eq:pon:dual:c}) under current link lengths computed according to Section \ref{sec:kernel}.

If the min-cost value is larger than one, the request is rejected.
Otherwise, the request is accepted, and the entire flow $d_r$ of request $r$ is routed along the min-cost SR-path.

After the flow is sent, the flow value and the link length at each link along the path $P_r^{k^*}$ are updated as follows:

1) Update the flow value $X_r^{{k}}$ as
\[X_r^{k^*} \leftarrow 1,\]
which implies $X_r^{k} = 0,\forall k \ne {k^*}$ according to constraint (\ref{eq:pon:f}).

2) Update the link lengths $l_e$ as
\[{l_e} \leftarrow {l_e}\left( {1 + \frac{{g_r^{k^*}(e){d_r}}}{{{c_e}}}} \right){\rm{ + }}\frac{\phi }{{Qn}}\frac{{g_r^{k^*}(e){d_r}}}{{{c_e}}},\phi >0\]

Parameter $\phi$ is designed to provide a tradeoff between the competitiveness of the proposed online algorithm and the degree of violating the capacity constraint in the primal problem. That is, a smaller $\phi$ leads to larger network throughput as well as a larger degree of violation on the link capacity \cite{anu_unicast}.

%
%
	
	\textbf{Theorem 2:}
	Algorithm \ref{alg_online} is an all-or-nothing, non-preemptive, monotone, and $\left\{ {O\left( 1 \right),O\left( {\log n} \right)} \right\}$-competitive,  more specifically $\left\{ {{{1 + \phi }},\log \left( {\frac{{BQn}}{\phi } + 1} \right)} \right\}$-competitive, online algorithm.
	In other words, the routing flow is $O(1)$-competitive and it violates the link capacity constraints by $O(\log n)$.
	
	\textit{Proof:}
	See Appendix.
	\qed
	
\section{Min-Cost SR-Path Computation}\label{sec:kernel}



The key steps in the FPTAS and the online algorithm all involve the computation of the min-cost SR-path for a request where the length of a link $e$ is the dual variable $l_e$.

In order to accelerate the FPTAS, the auxiliary graph construction and the link lengths update are organized into independent algorithms \textit{auxiliary graph construction} and \textit{mincost computation module}, respectively.
In the FPTAS, the auxiliary graph construction needs to be executed only once in each iteration while the mincost computation module should be executed in every step.
However, it makes no difference for the online algorithm whether the two algorithms are independent because both of them are executed once in every iteration.

\begin{figure}[!th]
	\centering
	\includegraphics[angle=0, width=0.47\textwidth]{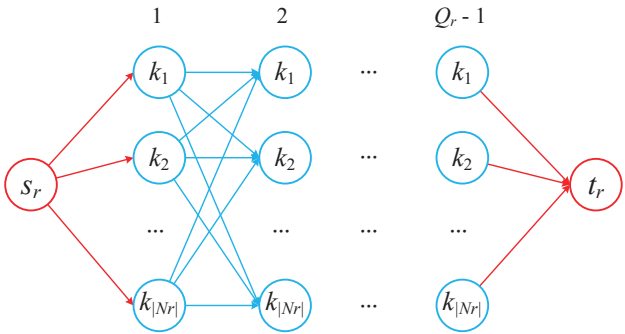}
	\caption{Auxiliary graph ${\widetilde G_r} = ({\widetilde V_r},{\widetilde E_r})$ for request $r$.}
	\label{fg:agc}
\end{figure}

In the \textit{auxiliary graph construction}, denote the \textit{auxiliary graph} by ${\widetilde G_r} = ({\widetilde V_r},{\widetilde E_r})$.
There are two end nodes corresponding to $s_r$ and $t_r$ for the current request $r$, and $Q_r-1$ layers of SR-nodes $N_r$.
There are links connecting $s_r$ to all the SR-nodes in the first layer, 
from each SR-node in the first layer to each SR-node in the second layer, etc,
This process is repeated until all SR-nodes of the last layer are connected to $t_r$, as shown in Fig.~\ref{fg:agc}.
The node set composed of all the possibly involved SR-nodes for a given request is called the \textit{Middlebox Fabric} (MF).

In the \textit{mincost computation module}, the main processes are as follows:

1) Execute an \textit{all-pair-flow-splitting-cost} (APFSC) computation to get the total costs between all node pairs. 
In particular, if the shortest path routing is employed in $f$, the APFSC computation reduces to an all-pair-shortest-path computation.
This, of course, can simplify the implementaiton and accelerate the algorithm.
We are interested in the paths and corresponding SR-nodes that are relevant for request $r$.

2) Update the link lengths of $\widetilde G_r$ to the total cost between the two nodes the link connects. 

3) Compute the shortest path in the auxiliary graph between $s_r$ and $t_r$. 
This determines the optimal segment list (SR-node list) $k$.

\begin{algorithm}[!ht]
	\caption{Auxiliary Graph Construction}\label{alg_construction}
	\begin{algorithmic}
		\STATE \textbf{Input:} $G$; $r$; $N_r$; $Q_r$
		\STATE Construct an auxiliary graph ${{\widetilde{G}}_{r}}=({{\widetilde{V}}_{r}},{{\widetilde{E}}_{r}})$ for request $r$.
		\STATE \textbf{Output:} ${\widetilde G_r}$
	\end{algorithmic}
\end{algorithm}
\begin{algorithm}[!ht]
	\caption{Mincost Computation Module}\label{alg_update}
	\begin{algorithmic}
		\STATE \textbf{Input:} ${\widetilde G_r}$; $l$
		\STATE Calculate the APFSC under the dual link length system $l$:
		\STATE \quad $c(u,v) = \sum\limits_e {{l_e}{f_{uv}}(e)} ,\forall u,v \in {N_r} \cup \{ {s_r},{t_r}\} ,\forall r$
		\STATE Update link lengths for ${\widetilde G_r}$: 
		\STATE \quad ${{\widetilde{l}}_{uv}}=c(u,v),\forall u,v\in \widetilde{V_r}$
		\STATE Compute the shortest path from $s_r$ to $t_r$ in ${{\widetilde{G}}_{r}}$ under the link length system $\widetilde{l}$.
		\STATE The length of the shortest path is:
		\STATE \quad $L:= \mathop {\min }\limits_{k \in {K_r}} \sum\limits_e {g_r^k(e){l_e}}$
		\STATE The optimal SR-node list is:
		\STATE \quad $k^*:= \arg \mathop {\min L}\limits_{k \in {K_r}}$
		\STATE (Note that the shortest path in ${{\widetilde{G}}_{r}}$ is exactly the min-cost SR-path in $G$ for request $r$.)
		\STATE \textbf{Output:} $k^*$; $g_{r}^{k^*}(e)$; $L$
	\end{algorithmic}
\end{algorithm}
Although the MF in Fig.~\ref{fg:agc} looks similar to that in \cite{steering}, the computation of end-to-end paths is entirely different.
For the algorithm in \cite{steering}, only the dual link weights are used to perform an all-pair shortest path computation in each iteration.
For our algorithms, the primal link weights are used to generate physical routing paths while the dual link lengths to guide flow allocation on the generated paths.

In practice, for the network operator, the MF in Fig.~\ref{fg:agc} can be automatically or even manually adapted to specific network topologies to steer the flow on a more desirable routing path while greatly reducing resource overheads.

\section{Simulation Results}
\subsection{Simulation Settings}
\begin{figure}[!ht]
	\centering
	\subfloat[Case I]{\includegraphics[width=0.33\textwidth]{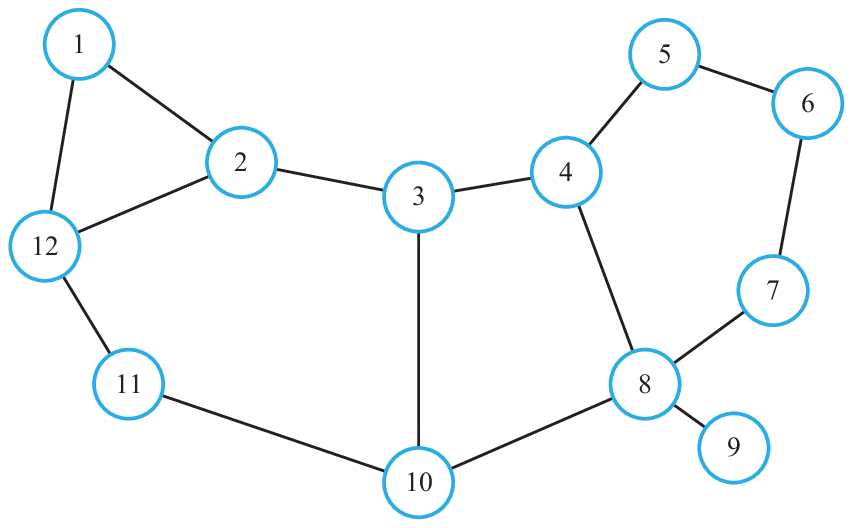}%
		\label{fg:topo:abilene}}
	\hfil
	\subfloat[Case I]{\includegraphics[width=0.485\textwidth]{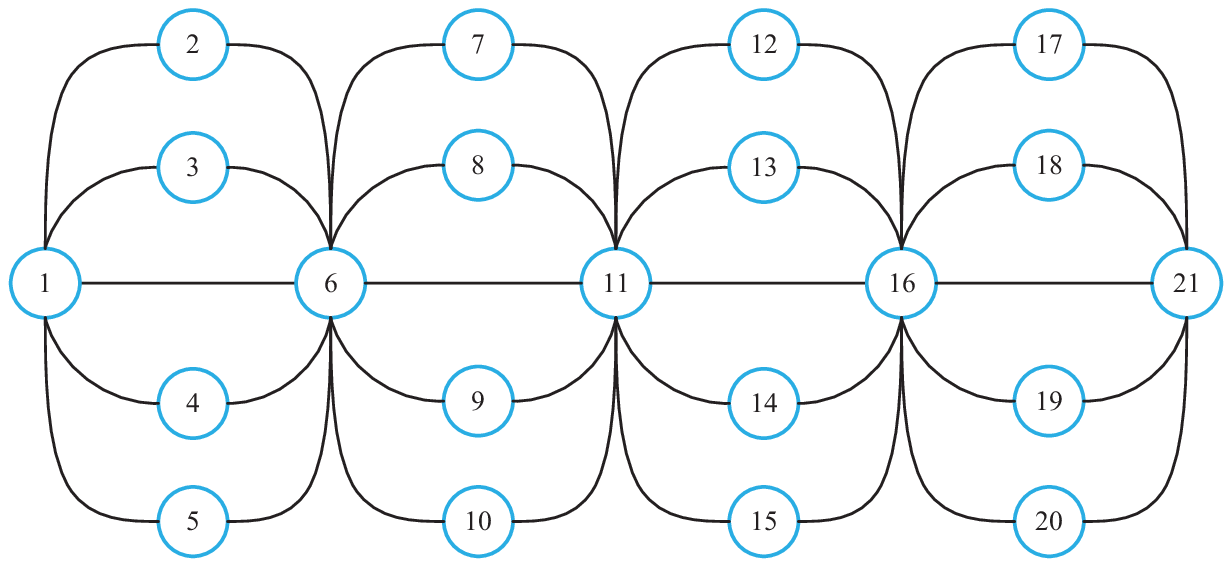}%
		\label{fg:topo:nsr}}
	\caption{Simulation networks. (a) Abilene network, 12 nodes and 30 links. Blackbox test. (b) $Q$-SR network, 21 nodes and 36 links. Whitebox test.}
	\label{fg:topo}
\end{figure}

We use two typical networks to evaluate the proposed solutions.
In the Abilene network shown in Fig.~\ref{fg:topo:abilene}, all the 30 links are bidirectional and have equal capacities 100.
In the $Q$-SR network shown in  Fig.~\ref{fg:topo:nsr}, without loss of generality, we make all the 36 links unidirectional from node 1 to 21.
The simulation settings are summarized in Table~\ref{tb:setting}.
The volume is an rough estimation to the whole network capacity and is calculated as the sum of all the link capacities.
	\begin{table*}[!h]\small
	\begin{adjustwidth}{-1.0in}{-1.0in}
		\caption{\label{tb:setting}Simulation Settings}
		\centering
		\begin{tabular}{p{3.2cm}p{1.5cm}p{2.8cm}p{4.5cm}p{3.5cm}}
			\toprule
			\textbf{Strategy} 				& \textbf{Capacity} & \textbf{Volume} & \textbf{FPTAS} & \textbf{Online algorithm}\\
			\midrule
			\textbf{Abilene network} 	 & 100 				& $100\times30=3000$ 		 & 12 random node pairs; $d_r$=20 		& 100 random requests; $d_r$=5 \\
			\textbf{$Q$-SR network} & 100 		& $100\times36=3600$ 			& 1 node pair; $d_r$=100 					& 100 requests; $d_r$=5\\
			\bottomrule
		\end{tabular}
	\end{adjustwidth}
\end{table*}
\begin{figure*}[!ht]
	\centering
	\subfloat[Case I]{\includegraphics[width=0.43\textwidth]{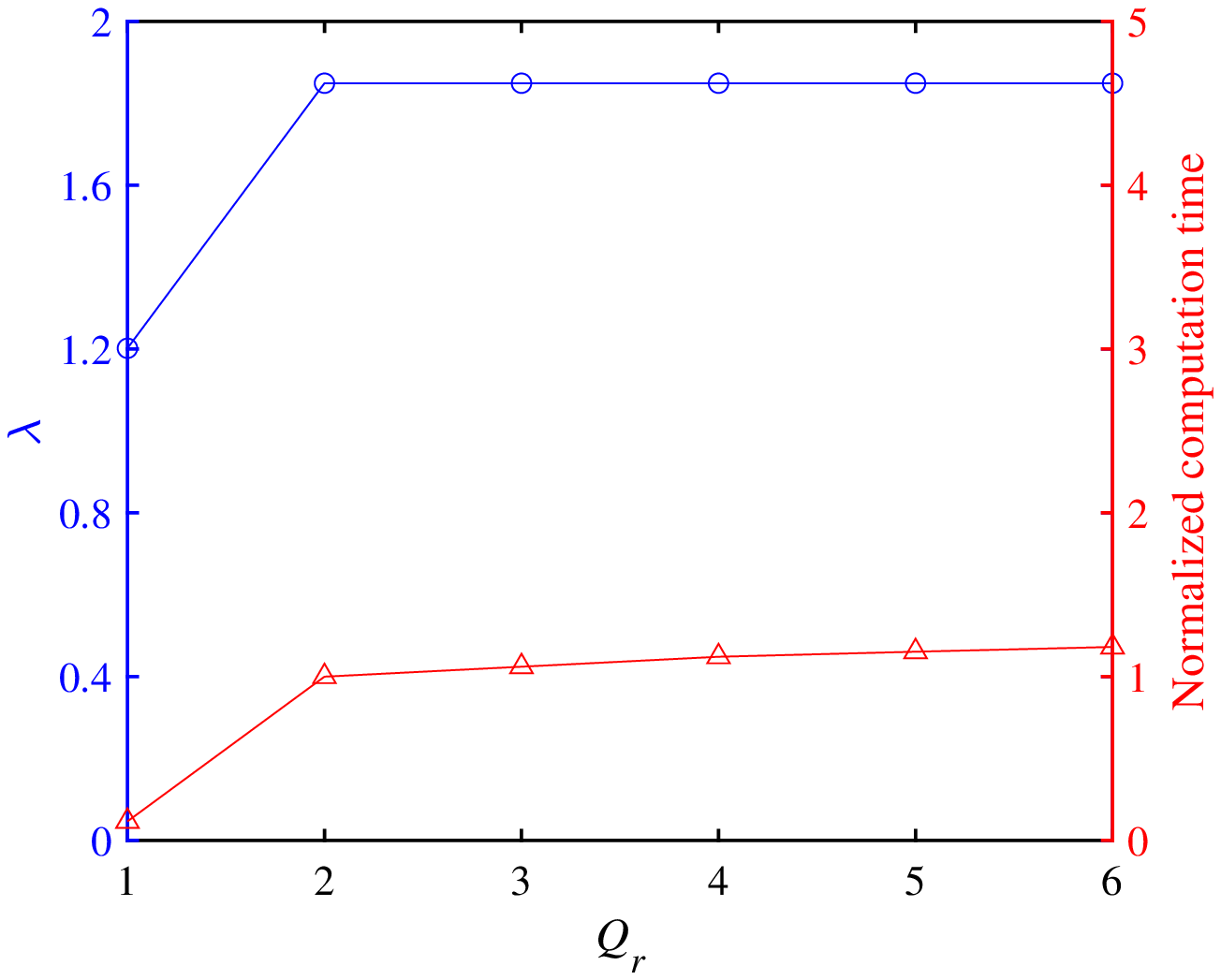}%
		\label{fig:fptas:qr:abi}}
	\hfil
	\subfloat[Case I]{\includegraphics[width=0.43\textwidth]{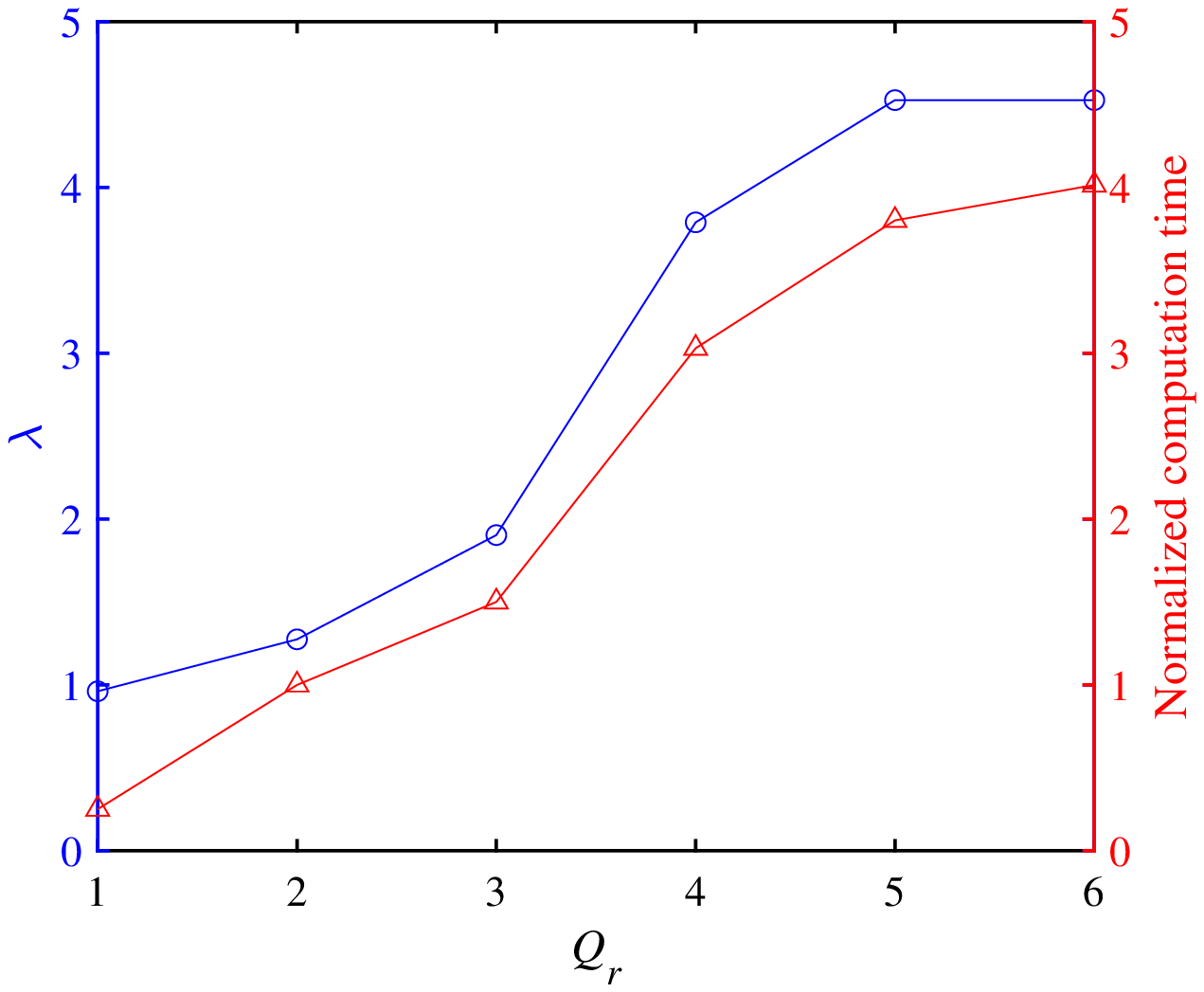}%
		\label{fig:fptas:qr:nsr}}
	\caption{FPTAS. $\lambda$ and normalized computation time versus $Q_r$. $\epsilon=0.1$. (a) Abilene network. (b) $Q$-SR network. }
	\label{fig:fptas:qr}
\end{figure*}
\begin{figure*}[!ht]
	\centering
	\subfloat[Case I]{\includegraphics[width=0.43\textwidth]{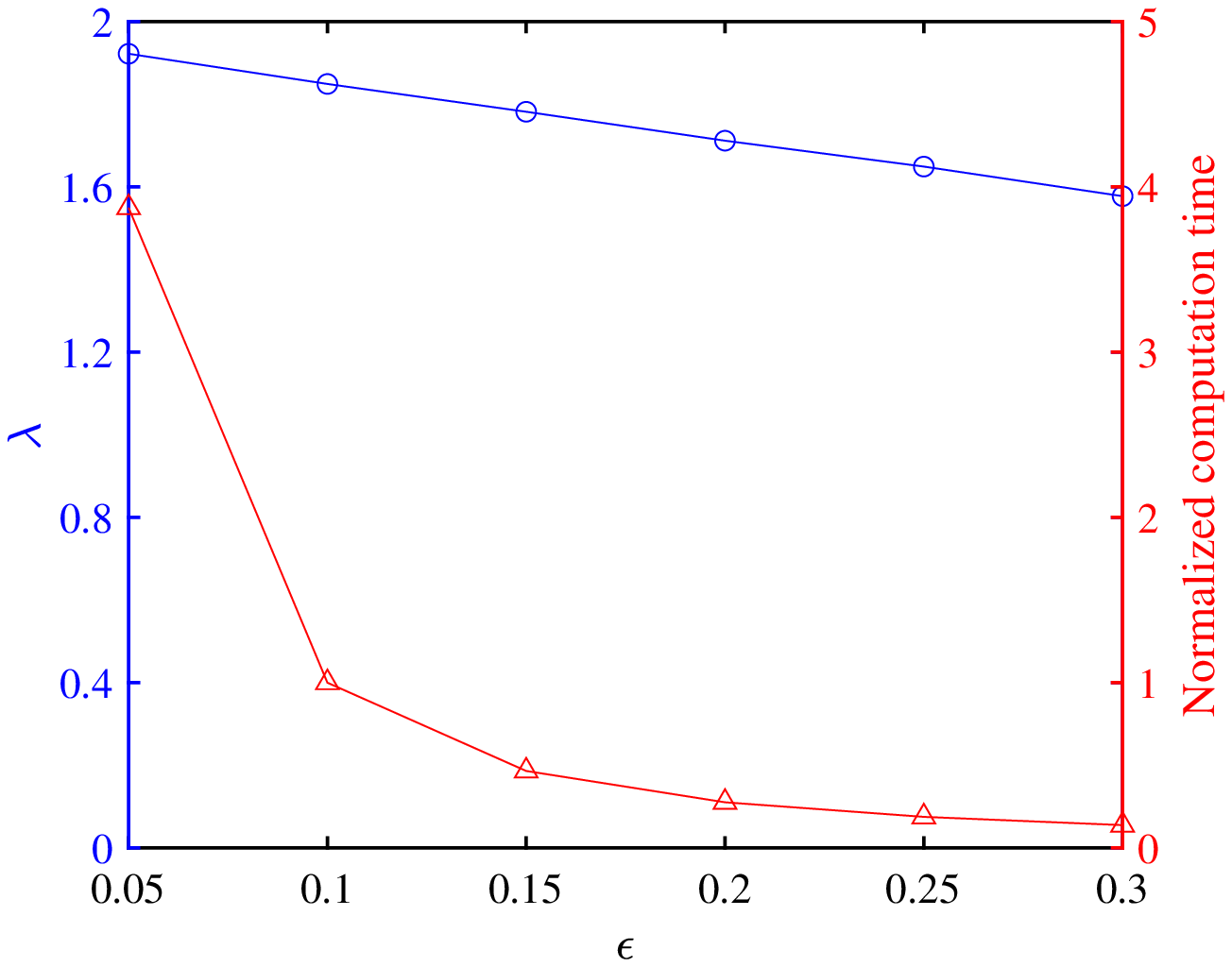}%
		\label{fig:fptas:epsilon:abi}}
	\hfil
	\subfloat[Case I]{\includegraphics[width=0.43\textwidth]{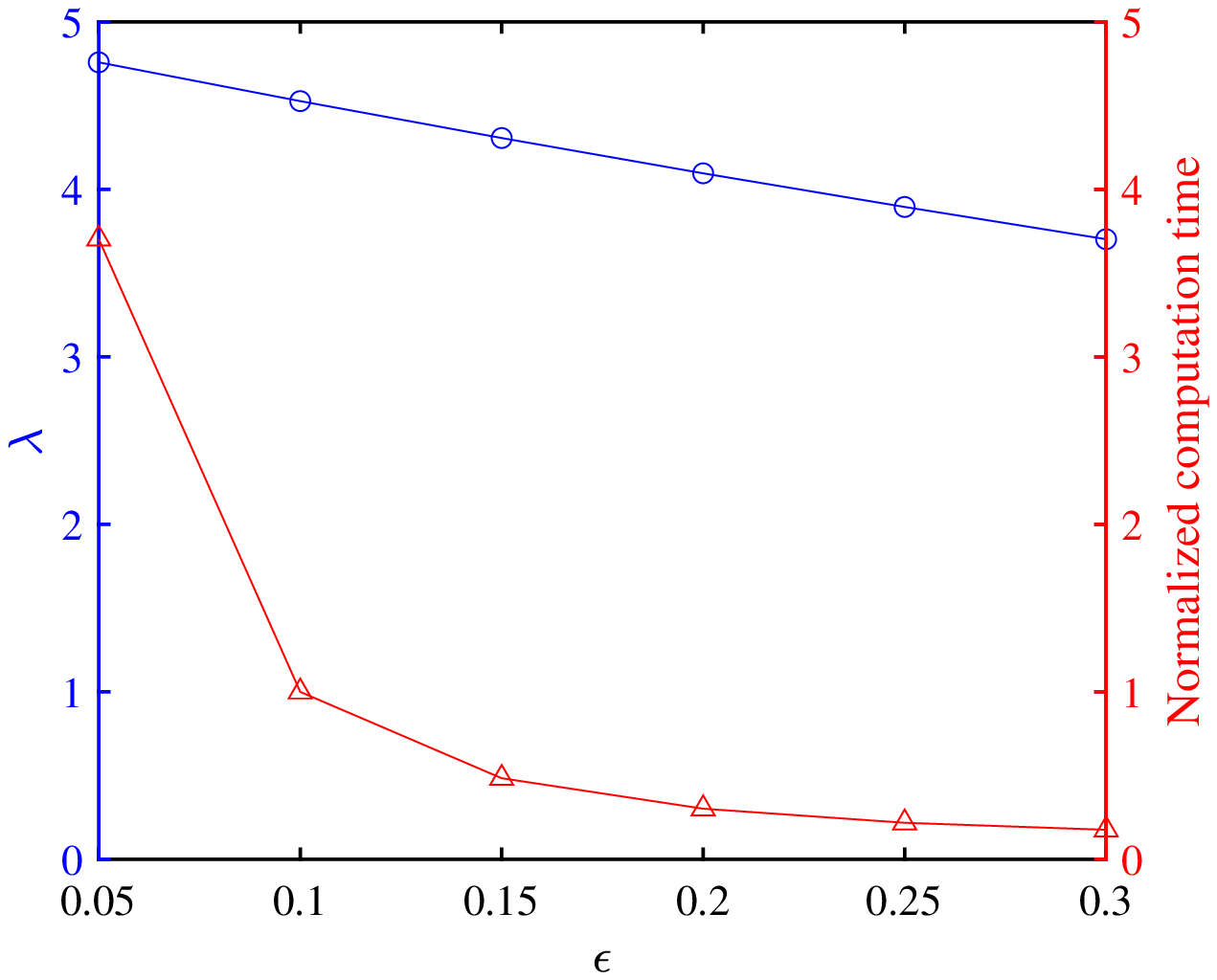}%
		\label{fig:fptas:epsilon:nsr}}
	\caption{FPTAS. $\lambda$ and normalized computation time versus $\epsilon$. (a) Abilene network. $Q_r=2$. (b) $Q$-SR network. $Q_r=5$. }
	\label{fig:fptas:epsilon}
\end{figure*}

The experiments on the Abilene network can be seen as \textit{blackbox} tests.
It shows the overall performance in realistic networks, especially the Internet backbone.
The experiments on the $Q$-SR network can be seen as \textit{whitebox} tests.
The reasons why we devise such a network are two-fold.
First, it essencially reflects the hierarchical characteristics of current multi-domain Internet \cite{Nestor,multi_domain}.
Specifically, this network simulates a real inter-domain network.
There exist multiple available paths between a node pair within a domain and
different domains are connected by edge devices which may be performance bottlenecks.
Second, the optimal throughput of Q-SR is hard to obtained by conventional methods.
For instance, it is nearly impossible to compute a 5-SR setting using a general LP solver even for a network with such size,
while we can easily make an estimation from this highly structured topology.
In this way, the traffic distribution becomes tractable and we are able to evaluate to what extent the proposed algorithms can approximate the optimum.


In the following, we conduct the simulations from two perspectives.
From the algorithimic perspective,
we want to validate and analyze the effects of parameters $\epsilon$ and $\phi$.
From the SR perspective,
since we have already parameterized the SR-node number, the segment number and the multipath policy for intra-segment routing in our model,
we will not validate all of the parameters or variables in this paper due to space limitation.
Instead, we are more interesting in whether $Q_r \ge 3$ would virtually influence the routing performance and resource consumption.
This is because lots of literature claim that it is unprofitable to set $Q_r \ge 3$ in real networks.

Generally speaking, a larger $N_r$ (and hence $Q_r$) will lead to a larger thoughput while accompanied by heavier computation overheads in the offline setting and severer bandwidth constraints violations in the online setting.
How to achieve a trade-off is closely relevant to the network topology and thus is not the focus of this paper.
Since we aim to highlight the computation efficiency of the proposed algorithms,
we let $N_r=N$ unless otherwise specified.

To evaluate the FPTAS, 
for the Abilene network, each of the 12 nodes randomly selects another node to send a request with size $d_r=20$;
for the $Q$-SR network, all the requests aggregate to one request from node 1 to 21 with size $d_r=100$.

To evaluate the online algorithm,
we randomly generate 100 requests with equal size $d_r=5$.
These requests enter the network one by one in a non-preemptive manner.
That is, once a request enters the network, it will stay for ever.
For the Abilene network, the traffic spreads across amost the whole network.
For the $Q$-SR network, the traffic is injected into the network at node 1 and is finally absorbed at node 21.

\subsection{FPTAS}
In this section, we validate the parameters $Q_r$ and $\epsilon$ in terms of the routing performance metric $\lambda$ and the computation cost metric.
The \textit{normalized computation time} is defined as the ratio of real computation time to the real computation time when $\epsilon=0.1$ (a commonly used setting in literature).
More precisely, the algorithms execute within a few seconds for all the instances considered in our simulation.

Fig.~\ref{fig:fptas:qr} shows how $Q_r$ influences the throughput as well as the computation overheads.
For the Abilene network, the throughput reaches the optimum when $Q_r=2$ and further increasing of $Q_r$ does not bring any improvent.
For the $Q$-SR network, the throughput gradually increases with $Q_r$ until it reaches the maximum 4.53 when $Q_r=5$.
It can be easily seen from Fig.~\ref{fg:topo:nsr} that the theoretical optimal throughput is $\lambda=5$ when $Q_r=5$.
All the 5 parallel paths, e.g. (1,2,6,7,11,12,16,17,21), from node 1 to 21 are fully utilized.
On the other hand, the optimum can only be achieved when $N_r=N$, more exactly $N_r=N-\{1,21\}$.
In fact, our algorithms can even support a rapid computation for a very large $Q_r$, say $Q_r=n-1$, with only small additional overheads than the $Q_r$ that just reaches the optimum. 

Considering that the Abilene network can reach a satisfactory throughput when $Q_r=2$ while $Q_r=5$ is the best setting for the $Q$-SR network,
we use these settings of $Q_r$ for the validation of $\epsilon$ and $\phi$.

Notably, the computation overheads also concern with the topologies.
The computation time increases almost simultaneously with the throughput in the $Q$-SR network, while there is only a slow increasing in the Abilene network.

Fig.~\ref{fig:fptas:epsilon} shows that the effects of $\epsilon$ are quite similar in the two networks.
When $\epsilon$ becomes smaller, $\lambda$ grows linearly while the computation time grows exponentially.
Obviously, $\epsilon=0.1$ is a good choice to reach a tradeoff between routing performance and computation cost.
For the $Q$-SR network, the setting $\epsilon=0.05$ leads to a throughput $\lambda=4.76$ when $Q_r=5$, which is fairly close to the optimum $\lambda=5$.

\begin{figure*}[!ht]
	\centering
	\subfloat[Case I]{\includegraphics[width=0.43\textwidth]{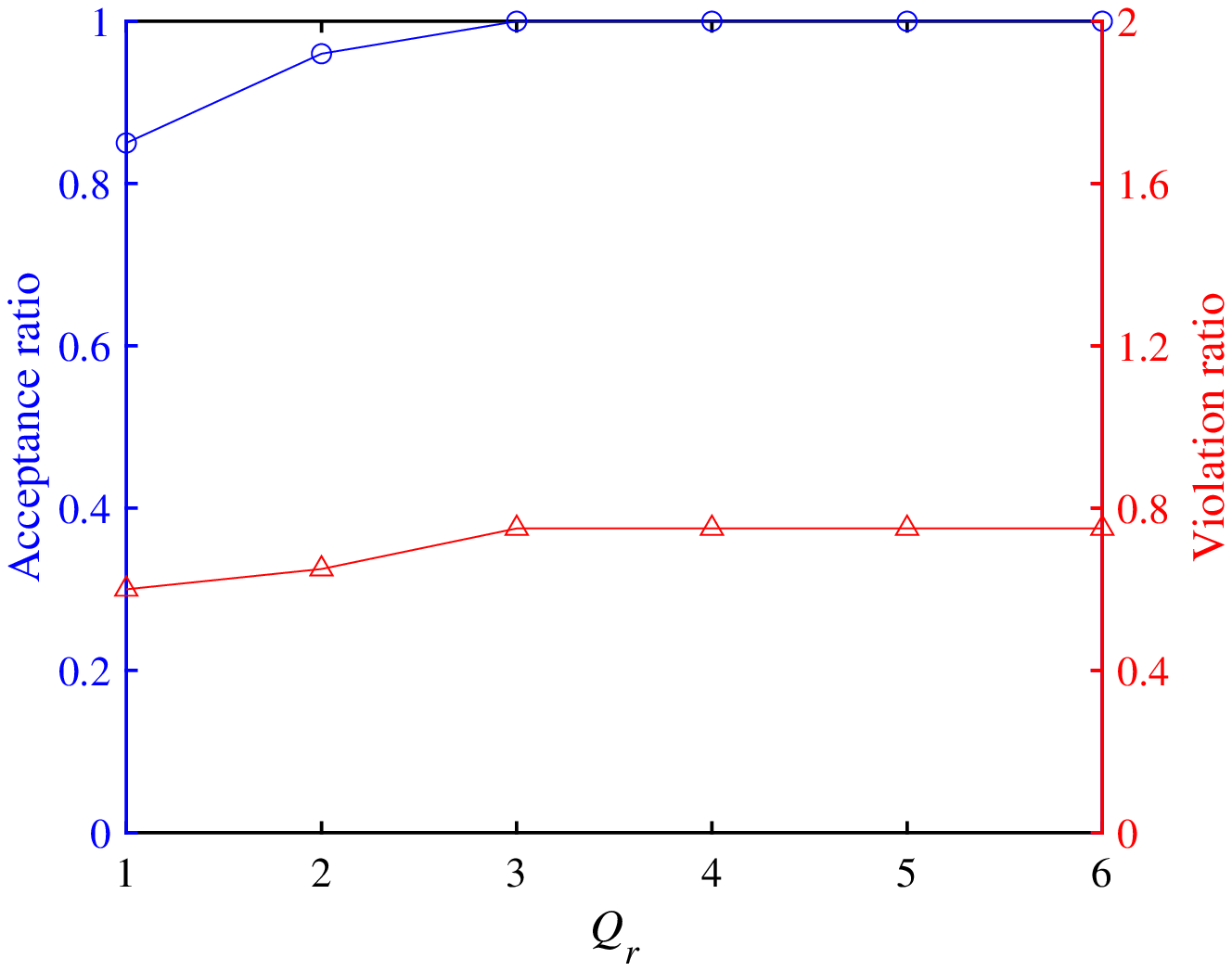}%
		\label{fig:online:qr:abi}}
	\hfil
	\subfloat[Case I]{\includegraphics[width=0.43\textwidth]{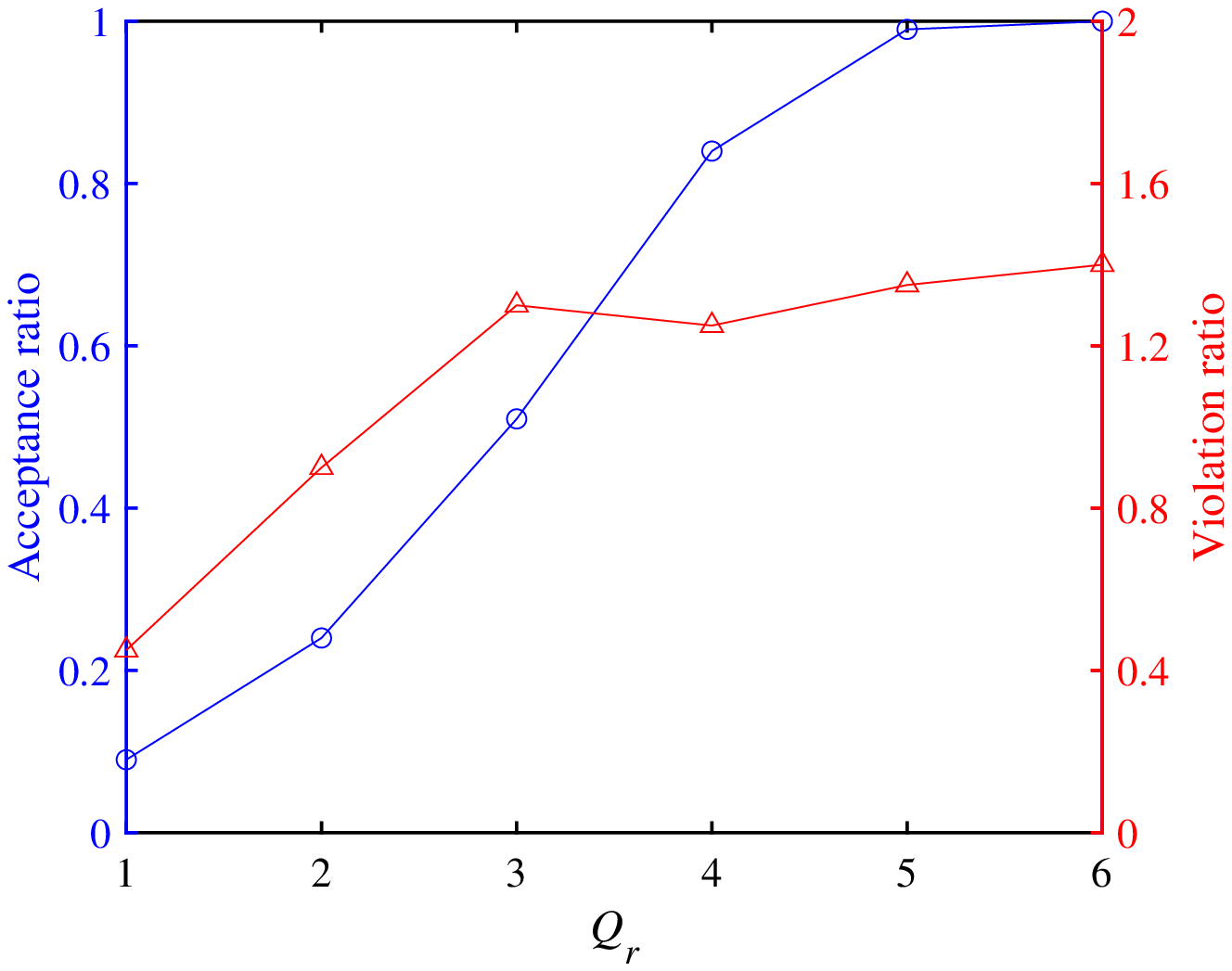}%
		\label{fig:online:qr:nsr}}
	\caption{Online algorithm. Acceptance ratio and violation ratio versus $Q_r$. $\phi=10$. (a) Abilene network. (b) $Q$-SR network.}
	\label{fig:online:qr}
\end{figure*}
\begin{figure*}[!ht]
	\centering
	\subfloat[Case I]{\includegraphics[width=0.43\textwidth]{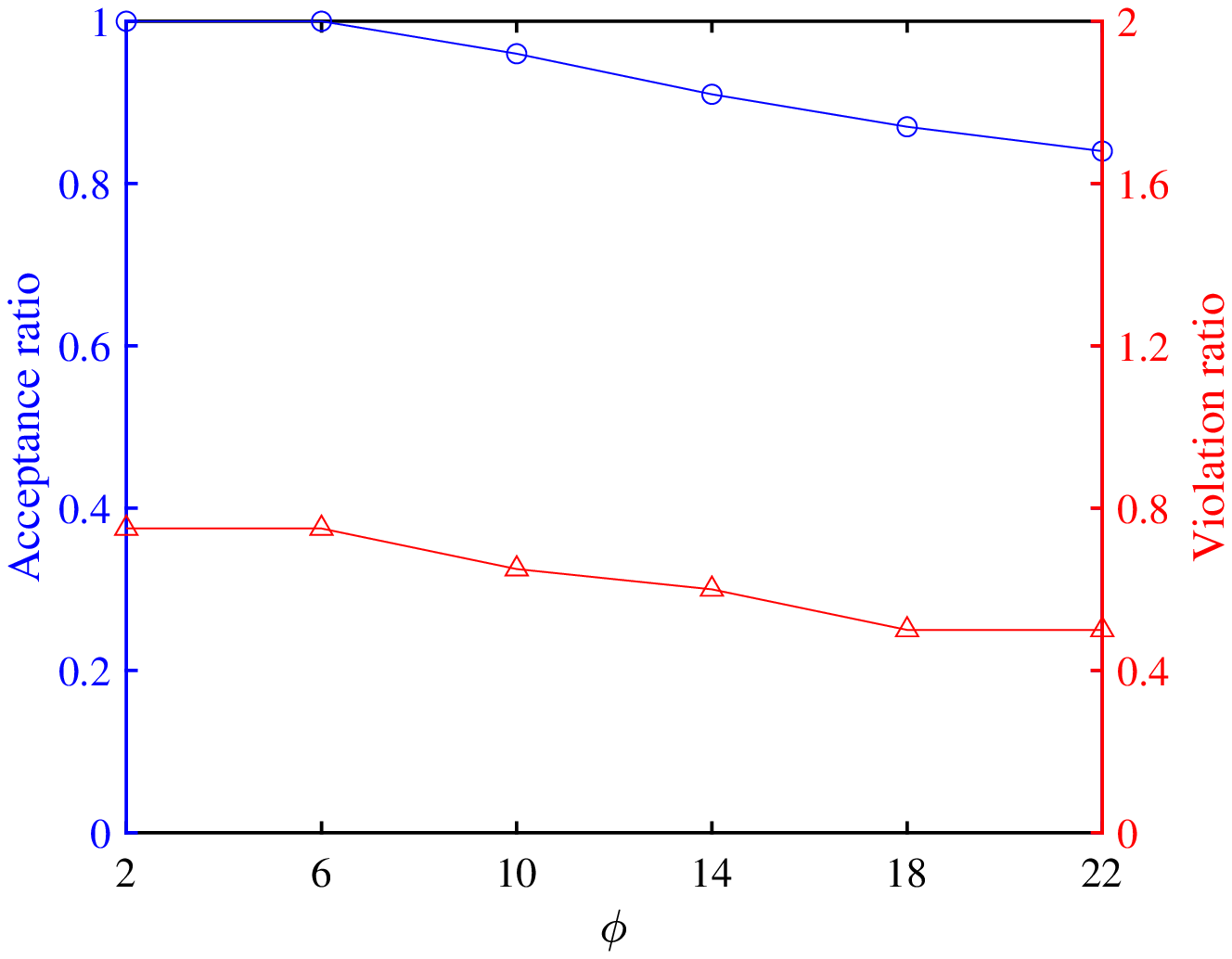}%
		\label{fig:online:phi:abi}}
	\hfil
	\subfloat[Case I]{\includegraphics[width=0.43\textwidth]{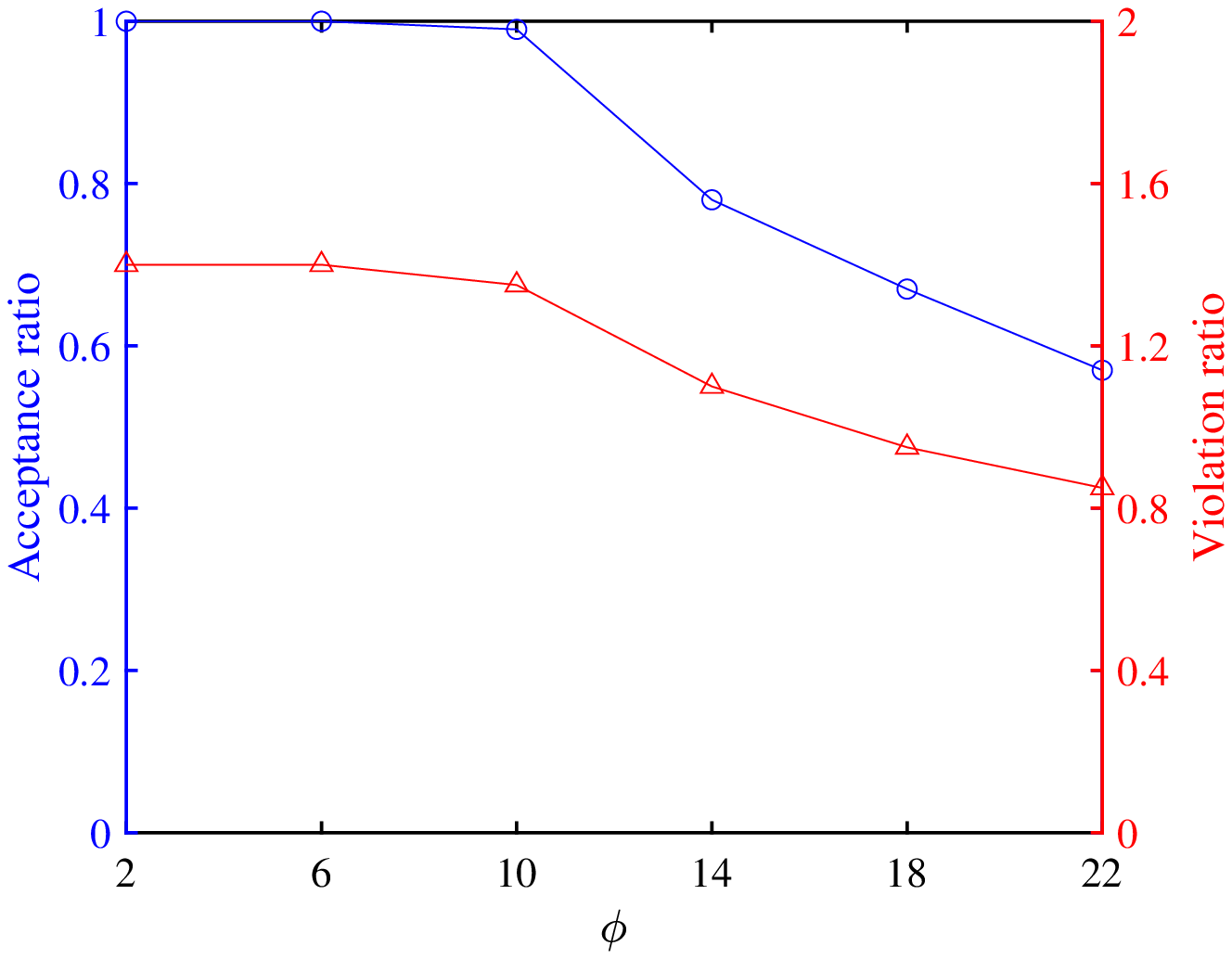}%
		\label{fig:online:phi:nsr}}
	\caption{Online algorithm. Acceptance ratio and violation ratio versus $\phi$. (a) Abilene network. $Q_r=2$. (b) $Q$-SR network. $Q_r=5$.}
	\label{fig:online:phi}
\end{figure*}

\subsection{Online Algorithm}

In this section, we validate the parameters $Q_r$ and $\phi$ in terms of the routing performance metric \textit{acceptance ratio} and the resource cost metric \textit{violation ratio}.
The \textit{acceptance ratio} is defined as the ratio of accepted number of requests to the total request number.
The \textit{violation ratio} is defined as the maximum ratio of the real flow amount on a link to its capacity over all links.

Similar to the offline scenario, as seen from Fig.~\ref{fig:online:qr}, $Q_r$ has significant influences on the routing performance as well as the resource consumption.
However, unlike the offline scenario, approaching the optimum needs an even larger $Q_r$ in the online setting.
For the Abilene network, the acceptance ratio reaches the optimum when $Q_r=3$.
For the $Q$-SR network, the acceptance ratio gradually increases with $Q_r$ until reaches the optimum when $Q_r=6$.

As for the violation ratio, the two networks have a similar trend.
The violation ratio gradually increases and reaches almost stable when $Q_r$ surpasses some value.

As shown in Fig.~\ref{fig:online:phi}, the effects of $\phi$ are also quite similar in the two networks.
When $\phi$ becomes smaller, $\lambda$ grows linearly while the computation time grows exponentially.

Similar to the effects of $\epsilon$ imposed on the FPTAS, there is also a trade-off between routing performance and resource consumption when choosing an appropriate $\phi$.
It is virtually meaningless to compare the performance between the two neworks, because the acceptance ratio can be raised by reducing the request sizes or enlarging the link capacities.

%

%

%
%
\section{Discussion}
The proposed $Q$-SR framework is flexible enough to be extended in the following ways. 

\textbf{Segment multicast:} The proposed framework as a whole can be extended to a novel routing paradigm \textit{segment multicast}.
By doing this, the SR-path becomes a pseudo directed steiner tree \cite{anu_multicast}.
Since the computation of a directed steiner tree is NP-hard, we can invoke an approximation algorithm in the min-cost SR-path computation model.
Of course, this may introduce some implementation issues and protocol overheads on encoding a multicast tree to packet header in a source routing manner. 

\textbf{Intra-segment routing:} The intra-segment routing should be fully investigated, including link weights setting and intra-segment routing policy \cite{trimponias2019node}.
For instance, the intra-segment routing module can be replaced with other link-state routing policies or even a centralized min-cost MCF routing module.
In the current SR architecture, the intra-segment routing is fixed and the only optimization space left lies in the inter-segment routing.

\textbf{SR-node selection and placement:} The topology-adaptive and traffic-aware SR-node selection and placement methods should be further considered.
Specifically, each layer of SR-nodes in the MF structure shown in Fig.~\ref{fg:agc} can be independently specified for each request. 
In the worst case, e.g. the $Q$-SR network, only if all the intermediate nodes are employed will the throughput be maximized.
Therefore, how to improve the routing performance while keeping as small overall costs as possible is also a major challenge.

\textbf{Combining with SFC:}
How to combine the research methods and results of SR and SFC, just as indicated in Section \ref{sec:relatedwork}, has both theory value and practical significance. 
From the standpoint of network operator, for instance, there is a strong need to incorporate a realistic SR-node cost model to the framework, while this may have been well solved in the context of SFC \cite{anu_unicast,anu_multicast}.

\section{Conclusion}
In this paper, 
we propose a flexible $Q$-SR model and its formulation where segment number, SR-node number, intra-segment routing policy are all parameterized. 
The model leads to a highly extensible framework to design and evaluate algorithms that can be adapted to various network topologies and traffic matrices.
For both offline and online settings, we develop primal-dual algorithms with provable worst case performance bounds.
The advantage of computation efficiency of the algorithms over existing methods is so great that it enables quantitative evaluation of various SR parameters and algorithmic parameters on various types of network topologies.

\section*{Appendix}
\subsection{Proofs for Algorithm~\ref{alg_fptas}}\label{sec:appendix:fptas}
\textbf{Lemma 1:} When the FPTAS terminates, the primal solution needs to be scaled by a factor of at most 
${\log _{1 + \epsilon }}\frac{1}{\delta }$ to ensure primal feasibility (i.e., satisfying link capacity constraints).

\textit{Proof:}
Serialize all the steps of all the iterations of all the phases into $l$ steps.
Define the \textit{flow scaling factor} of link $e$ as:

\[\kappa : = \sum\limits_{i = 1}^l {\frac{{g_{r(i)}^{k(i)}(e){\Delta _{(i)}}}}{{{c_e}}}}  \]

According to the update rule of $l_e$, we have:
\[l_e^{\rho  - 1,K} = \frac{\delta }{{{c_e}}}\prod\limits_{i = 1}^l {\left( {1 + \epsilon \frac{{g_{r(i)}^{k(i)}(e){\Delta _{(i)}}}}{{{c_e}}}} \right)} \]

%

Using the \textit{Taylor Formula}, the inequality ${(1 + x)^a} \le 1 + ax,\forall x,\forall a \in [0,1]$ holds.
Setting $x = \epsilon$ and $a = {\frac{{g_{r(i)}^{k(i)}(e){\Delta _{(i)}}}}{{{c_e}}}} \le 1$, we have:
\begin{align*}
1 > l_e^{\rho  - 1,K} &\ge \delta \prod\limits_{i = 1}^l {{\left( {1 + \epsilon } \right)}^{\frac{{g_{r(i)}^{k(i)}(e){\Delta _{(i)}}}}{{c_e}}}} \nonumber\\
&{ = }\delta {\left( {1 + \epsilon } \right)^{\sum\limits_{i = 1}^l{\frac{{g_{r(i)}^{k(i)}(e){\Delta _{(i)}}}}{{{c_e}}}} }} \nonumber\\
&{ = }\delta {\left( {1 + \epsilon } \right)^\kappa } \nonumber
\end{align*}

Hence, the lemma is proven.
\[\kappa  < {\log _{1 + \epsilon }}\frac{1}{\delta }\]

\qed

\textbf{Lemma 2:}
At the end of $\rho$ phases in the FPTAS, we have
\[\frac{\beta }{{\rho  - 1}} \le \frac{\epsilon }{{(1 - \epsilon )\ln \frac{{1 - \epsilon }}{{n\delta }}}}\]

\textit{Proof:}
Define
\[{\rm{mincos}}{{\rm{t}}_r}({l^{i,r,s}}) = \mathop {\min }\limits_{k \in {K_r}} \sum\limits_e {g_r^k(e)l_e^{i,r,s}} \]
\begin{align*}
D({l^{i,r,s}})
&= \sum\limits_e {l_e^{i,r,s}{c_e}} \nonumber\\
&{ = }D({l^{i,r,s - 1}}) + \epsilon {\Delta ^{i,r,s}}\sum\limits_e {g_r^k(e)l_e^{i,r,s - 1}}  \nonumber\\
&{ = }D({l^{i,r,s - 1}}) + \epsilon {\Delta ^{i,r,s}}{\rm{mincos}}{{\rm{t}}_r}({l^{i,r,s - 1}}) \nonumber\\
&{ \le }D({l^{i,r,s - 1}}) + \epsilon {\Delta ^{i,r,s}}{\rm{mincos}}{{\rm{t}}_r}({l^{i,r,s}}) \nonumber
\end{align*}

\[D({l^{i,r}}) \le D({l^{i,r,0}}) + \epsilon {d_r}{\rm{mincos}}{{\rm{t}}_r}({l^{i,r}})\]

Define
\[\alpha (l) = \sum\limits_{r = 1}^{|R|} {{d_r}{\rm{mincos}}{{\rm{t}}_r}(l)} \]

We now sum over all iterations during phase $i$ to obtain:
\begin{align*}
D({l^{i,K}})
& {\le } D({l^{i,0}}) + \epsilon \sum\limits_{r = 1}^{|R|} {{d_r}{\rm{mincos}}{{\rm{t}}_r}({w^{i,K}})}  \nonumber\\
&{ = }D({l^{i,0}}) + \epsilon \alpha ({l^{i,K}}) \nonumber\\
&{ = } D({l^{i - 1,K}}) + \epsilon \alpha ({l^{i,K}}) \nonumber
\end{align*}

Since $\beta  \le \frac{{D({l^{i,K}})}}{{\alpha ({l^{i,K}})}}$, we have:
\[D({l^{i,K}}) \le \frac{{D({l^{i - 1,K}})}}{{1 - \epsilon /\beta }}\]

Using the initial value $D({l^{1,0}}) = m\delta $, we have for $i \ge 1$
\[D({l^{i,K}}) \le \frac{{m\delta }}{{1 - \epsilon }}{e^{\frac{{\epsilon (i  - 1)}}{{\beta (1 - \epsilon )}}}}\]

The last step uses the assumption that $\beta \ge 1$. 
The procedure stops at the first phase $\rho$ for which
\[1 \le D({l^{\rho ,K}}) \le \frac{{m\delta }}{{1 - \epsilon }}{e^{\frac{{\epsilon (\rho  - 1)}}{{\beta (1 - \epsilon )}}}}\]
which implies that
\[\frac{\beta }{{\rho  - 1}} \le \frac{\epsilon }{{(1 - \epsilon )\ln \frac{{1 - \epsilon }}{{m\delta }}}}\]

\qed

\textbf{Proof of Theorem 1:} 
The analysis of the algorithm proceeds similar to \cite{fptas2007}. 

\textit{1) Approximation ratio:}
Let $\gamma$ represent the ratio of the dual to the primal solution. Then we have
\[\frac{\mathcal{D}}{\mathcal{P}}:=\gamma <\frac{\beta }{\rho -1}{{\log }_{1+\epsilon }}\frac{1}{\delta }\]

Substituting the bound on $\frac{\beta }{\rho -1}$ from Lemma 2, we have
\[\gamma <\frac{\epsilon {{\log }_{1+\epsilon }}\frac{1}{\delta }}{(1-\epsilon )\ln \frac{1-\epsilon }{m\delta }}=\frac{\epsilon }{(1-\epsilon )\ln (1-\epsilon )}\frac{\ln \frac{1}{\delta }}{\ln \frac{1-\epsilon }{m\delta }}\]

Setting $\delta ={{\left( \frac{1-\epsilon }{m} \right)}^{\frac{1}{\epsilon }}}$ leads to $\gamma \le {{(1-\epsilon )}^{-3}}$.

Equating the desired approximation factor $(1+\zeta )$ to this ratio and solving for $\epsilon$, we get the value of $\epsilon$
stated in the theorem.

\textit{2) Running time:}
Using weak-duality from linear programming theory, we have
\[1\le \gamma <\frac{\beta }{\rho -1}{{\log }_{1+\epsilon }}\frac{1}{\delta }\]
Then the number of phases $\rho$ is upper bounded by
\[\rho =\left\lceil \frac{\beta }{\epsilon }{{\log }_{1+\epsilon }}\frac{m}{1-\epsilon } \right\rceil \]

Note that the number of phases derived above is under the assumption $\beta  \le 2$.
The case $\beta > 2$ can be recast as $1 \le \beta  \le 2$
by scaling the link capacities and/or request sizes using the same technique in Section 5.3 of \cite{fptas2007}. 
Then, the number of phases is at most $2\rho\log |R|$.
We omit the details here.

Since each link length has an initial value of $\frac{\delta}{c_e}$ and a final length less than $\frac{1+\epsilon}{c_e}$, 
the number of steps exceeds the number of iterations by at most $m{{\log }_{1+\epsilon }}\frac{1+\epsilon }{\delta}$.
Considering that each phase contains $|R|$ iterations, the total number of steps is at most
\[(2|R|\log |R| + m)\rho  \equiv O(\frac{{|R|\log |R|}}{\epsilon }{\log _{1 + \epsilon }}\frac{m}{{1 - \epsilon }})\]

Multiplying the above expression by $T_{\rm{SR}}$, i.e. the running time of each step, proves the theorem.
\qed

\subsection{Proofs for Algorithm~\ref{alg_online}}\label{sec:appendix:online}
\textbf{Proof of Theorem 2:} 
The online algorithm is by nature an approximation algorithm, and the performance guarantee can be proved in three steps as in \cite{bhatia2015sr}.

\textit{1) Dual feasibility:} We first show that the dual variables $l_e$ and $z_r$ generated in each step by the algorithm are feasible.

Let $k^*$ denote the intermediate node that minimizes
$\sum\limits_e {g_r^k(e){l_e}}$.

Setting 
$z_r = d_r\left( {1 - L } \right)$
makes
$z_r \ge {d_r}\left( {1 - \sum\limits_e {g_r^k(e){l_e}} } \right)$
hold for all SR-node lists.
The subsequent increase in $l_e$ will always maintain the inequality since $l_e$ does not change.

\textit{2) Competitive ratio:}
First, we give an upper bound of $\sum\nolimits_{e}{g_{r}^{k}(e)}$.
Suppose there are $q$ paths from node $u$ to $v$ and the flow amount on path $q$ is ${{\gamma }_{q}}$, then 
\[\sum\limits_{q}{{{\gamma }_{q}}}=1\]

Denote by $E_q$ the number of links of path $q$, then
\[\sum\limits_{e}{{{f}_{uv}}(e)}=\sum\limits_{q}{{{\gamma }_{q}}{{E}_{q}}}\le n\sum\limits_{q}{{{\gamma }_{q}}}=n\]

Thus, the SR-function for request $r$ is
\[\sum\limits_{e}{g_{r}^{k}(e)}=\sum\limits_{(u,v)\in k}{\sum\limits_{e}{{{f}_{uv}}(e)}}\le {{Q}_{r}}n\le Qn,\forall k\in {{K}_{r}}\]

During the step where request $r$ is accepted, the increase in the primal function is:
\[\Delta \mathcal{P}=d_r,\]
and the increase in the dual function is:
%
\begin{align*}
\Delta \mathcal{D} = {z_r}{\rm{ + }}\sum\limits_e {{c_e}\Delta {l_e}}  &= {d_r}\left[ {1 + \frac{{\phi \sum\nolimits_e {g_r^k(e)} }}{{Qn}}} \right] \nonumber\\
& \le {d_r}(1 + \phi ) \nonumber
\end{align*}

Therefore, the competitive ratio can be calculated as:
\[\frac{{\Delta \mathcal{D}}}{{\Delta \mathcal{P}}}{\rm{ = }}\frac{{{z_r}{\rm{ + }}\sum\nolimits_e {{c_e}\Delta {l_e}} }}{{{d_r}}} \ge 1 + \phi  \equiv O(1)\]

\textit{3) Primal feasibility:}  We now show that the solution is \textit{almost} primal feasible.


Denote the link price $l_e$ after request $r$ has been accepted and processed by $l(e,r)$,
and the utilization of link $e$ as 
\[\rho (e,r): = \frac{{F(e,r)}}{{{c_e}}}\]

First, we prove an lower bound of $l(e,r)$:
\[l(e,r) \ge \frac{\phi }{Qn}\left[ {{e^{\rho (e,r)}} - 1} \right]\]

We use the induction method.
According to the update rule of $l_e$, we have:
\begin{align*}
l(e,r) 
&\ge l(e,r - 1)\left( {1 + \frac{{g_r^{k^*}(e){d_r}}}{{{c_e}}}} \right){\rm{ + }}\frac{\phi }{Qn}\frac{{g_r^{k^*}(e){d_r}}}{{{c_e}}}\nonumber\\
& \ge \frac{\phi }{Qn}\left[ {{e^{\rho (e,r - 1)}} - 1} \right]\left( {1 + \frac{{g_r^{k^*}(e){d_r}}}{{{c_e}}}} \right){\rm{ + }}\frac{\phi }{Qn}\frac{{g_r^{k^*}(e){d_r}}}{{{c_e}}} \nonumber\\
& = \frac{\phi }{Qn}\left[ {{e^{\rho (e,r - 1)}}\left( {1 + \frac{{g_r^{k^*}(e){d_r}}}{{{c_e}}}} \right) - 1} \right] \nonumber\\
& \ge \frac{\phi }{Qn}\left[ {{e^{\rho (e,r)}} - 1} \right] \nonumber
\end{align*}


The last inequality follows from:
\[\left( {1 + \frac{{g_r^{k^*}(e){d_r}}}{{{c_e}}} } \right) \approx {e^{\frac{{g_r^{k^*}(e){d_r}}}{{{c_e}}}}}\]

and
\[\rho (e,r) = \rho (e,r - 1) + \frac{{g_r^{k^*}(e){d_r}}}{{{c_e}}}. \]


Second, we prove an upper bound of $l(e,r)$:
\[l(e,r) \le B\]

Denote ${g_{\min }} = \mathop {\min }\limits_e g_r^{{k^*}}(e)$. 
After request $r$ is accepted, the min-cost value $L \le 1$.
Then:
\[{g_{{\rm{min}}}}l(e,r - 1) \le L \le 1\]

According to the update rule of $l(e,r)$ and $\frac{{g_r^{k^*}(e){d_r}}}{{{c_e}}} \le 1$, we have:
\[l(e,r) \le \frac{1}{g_{{\rm{min}}}} \cdot {\rm{2 + }}\frac{\phi }{Qn} \cdot 1 = \frac{1}{g_{{\rm{min}}}}{\rm{ + }}\frac{\phi }{Qn}: = B\]

Combining the lower bound and the upper bound, we have:
\[\frac{F(e,r)}{c_e}:=\rho (e,r) \le \log \left( {\frac{{BQn}}{\phi } + 1} \right) \equiv O(\log n)\]

\qed

\end{document}